\documentclass{revtex4}
\usepackage{amsmath}
\usepackage{graphicx}

\begin{document}

\title{One- and two-dimensional solitons in $\mathcal{PT}$-symmetric systems
emulating the spin-orbit coupling}
\author{Hidetsugu Sakaguchi}
\affiliation{Department of Applied Science for Electronics and Materials,
Interdisciplinary Graduate School of Engineering Sciences, Kyushu
University, Kasuga, Fukuoka 816-8580, Japan}
\author{Boris A. Malomed}
\affiliation{Department of Physical Electronics, School of Electrical Engineering,
Faculty of Engineering, Tel Aviv University, Tel Aviv 69978, Israel\\
Laboratory of Nonlinear-Optical Informatics, ITMO University, St. Petersburg
197101, Russia}

\begin{abstract}
We introduce a two-dimensional (2D) system, which can be implemented in
dual-core planar optical couplers with the Kerr nonlinearity in its cores,
making it possible to blend effects of the $\mathcal{PT}$ symmetry,
represented by balanced linear gain and loss in the two cores, and
spin-orbit coupling (SOC), emulated by a spatially biased coupling between
the cores. Families of 1D and 2D solitons and their stability boundaries are
identified. In the 1D setting, the SOC leads, at first, to shrinkage of the
stability area for $\mathcal{PT}$-symmetric solitons, which is followed by
its rapid expansion. 2D solitons have their stability region too, in spite
of the simultaneous action of two major destabilizing factors, \textit{viz}%
., the collapse driven by the Kerr nonlinearity and a trend towards
spontaneous breakup of the gain-loss balance. In the limit of the SOC terms
dominating over the intrinsic diffraction, the 1D system gives rise to a new
model for gap solitons, which admits exact analytical solutions.

\end{abstract}

\maketitle

\section{Introduction}

The recent progress in the experimental and theoretical work with engineered
optical media has made it possible to emulate, by means of the optical-beam
propagation, a wide range of physical effects which were originally
predicted or experimentally discovered in other areas of physics. In many
cases, the optical emulation offers a possibility to study the effect in
question in a pure form, which is often too difficult in the original
setting. In other cases, specially designed optical configurations open a
way to demonstrate realizations of the effects in forms which are impossible
in the source systems. Belonging to this class of optically emulated
phenomena are the parity-time ($\mathcal{PT}$) symmetry \cite{Bender}, based
on the paraxial beam propagation in optical media with symmetrically placed
gain and loss elements, as predicted theoretically \cite{PT-optics} and
realized experimentally \cite{PT-optics-exp}; Anderson localization in
random photonic media \cite{Anderson}; and the photonic emulation of
topological insulators \cite{opt-top-insulator} and graphene \cite%
{opt-graphene}. A recent addition to this topic is a proposal to implement
the mechanism of the spin-orbit coupling (SOC), previously elaborated in
terms of Bose-Einstein condensates (BECs) \cite{SOC}, in dual-core optical
waveguides \cite{opt-SOC-2D,opt-SOC-1D}.

Basic schemes of ``grafting" new physical effects to optics
make use of linear beam-propagation regimes. However, the ubiquitous Kerr
nonlinearity of optical materials, as well as other types of the optical
nonlinear response, suggest one to consider implementation of the new
effects in a nonlinear form. A well-known example is provided by $\mathcal{PT%
}$-symmetric solitons, which have been studied in detail theoretically \cite%
{PT-sol-theory} and created in the experiment \cite{PT-sol-exp}. The optical
emulation of the SOC was also developed in the context of the nonlinear
propagation and formation of solitons \cite{opt-SOC-2D}.

As a further development in these directions, it may be interesting to
design optical settings which combine the above-mentioned effects, that
would be difficult to achieve in their original realizations. The objective
of the present work is to elaborate one- and two-dimensional (1D and 2D)
optical systems, based on dual-core waveguides, which make it possible to
blend the SOC and $\mathcal{PT}$-symmetry mechanisms. The use of couplers
for this purpose is natural, as they provide photonic platforms for the
emulation of both the SOC and $\mathcal{PT}$ symmetry in 1D \cite%
{opt-SOC-1D,Radik} and 2D \cite{opt-SOC-2D,Gena} geometries alike. As an
additional result of the analysis, a new 1D conservative model for
two-component gap solitons is produced, which admits exact solutions, and
demonstrates a nontrivial internal stability boundary in the soliton family.
New dynamical features are revealed by the analysis of the combined models,
such as a nonmonotonous dependence of the stability area of 1D $\mathcal{PT}$%
-symmetric solitons on the SOC strength, $\delta $: the area originally
shrinks but then strongly expands with the increase of $\delta $. Another
noteworthy finding is that the SOC terms stabilize 2D $\mathcal{PT}$%
-symmetric solitons under the action of the cubic self-attraction, in spite
of the simultaneous presence of \emph{two} mechanisms driving the
catastrophic instability in the 2D system: the possibility of the breakup of
the $\mathcal{PT}$ symmetry, i.e., failure of the gain-loss balance \cite%
{PT-optics,PT-sol-theory}, and the onset of the critical collapse induced by
the Kerr nonlinearity \cite{collapse} (the latter may also take place in the
presence of SOC \cite{Bilbao}). The latter result implies extension
of the SOC-induced stabilization of 2D solitons (\textit{semi-vortices} and
\textit{mixed modes}) in the free space with the cubic self-attraction, that
was demonstrated recently \cite{we}.

The paper is organized as follows. The basic model is introduced in Section
II. Results for 1D solitons and their stability are collected in Section
III. The new 1D model for gap solitons and its $\mathcal{PT}$-symmetric
version are presented in Section IV. Results for 2D solitons are reported in
Section V, and the paper is concluded by Section VI.

\section{The model}

The starting point is a model for the dual-core planar optical waveguide
which governs the spatiotemporal evolution of local amplitudes of the
electromagnetic waves in two cores, $U_{1}\left( x,t,z\right) $ and $%
U_{2}\left( x,t,z\right) $, under the action of the anomalous group-velocity
dispersion (GVD) and Kerr nonlinearity:
\begin{eqnarray}
i\left( U_{1}\right) _{z}+\frac{1}{2}\left[ \left( U_{1}\right) _{tt}+\left(
U_{1}\right) _{xx}\right] +U_{2}+\left\vert U_{1}\right\vert ^{2}U_{1} &=&0,
\label{1} \\
i\left( U_{2}\right) _{z}+\frac{1}{2}\left[ \left( U_{2}\right) _{tt}+\left(
U_{2}\right) _{xx}\right] +U_{1}+\left\vert U_{2}\right\vert ^{2}U_{2} &=&0.
\label{2}
\end{eqnarray}%
In these coupled 2D nonlinear Schr\"{o}dinger equations (NLSEs) $z$ is the
propagation distance, $x$ the transverse coordinate, $t$ the reduced
temporal variable, the second $x$ derivatives account for the paraxial
diffraction, while the coupling coefficient, together with the GVD and Kerr
coefficients, are scaled to be $1$. The 1D\ reduction of Eqs. (\ref{1}) and (%
\ref{2}) amounts to the commonly known 1D model for dual-core optical fibers
or the spatial propagation in the twin-core planar waveguides \cite{coupler}.

By itself, the system of Eqs. (\ref{1}) and (\ref{2}) gives rise solely to
unstable 2D solitons, due to the occurrence of the collapse in the same
setting. The stabilization of the spatiotemporal solitons against the
collapse may be provided by the optical counterpart of the SOC, which
amounts to taking into account the temporal dispersion of the coupling
coefficient \cite{opt-SOC-2D}, represented by terms $\delta \left(
U_{2}\right) _{t}$ and $\delta \left( U_{1}\right) _{t}$ to be added to Eqs.
(\ref{1}) and (\ref{2}), respectively, with a real dispersion coefficient $%
\delta $ \cite{Chiang}. The accordingly modified coupled equations, which
also include the balanced gain and loss terms, with real coefficient $\Gamma
>0$ \cite{Radik}, that introduce the $\mathcal{PT}$ symmetry, take the form
of
\begin{eqnarray}
i\left( U_{1}\right) _{z}+\frac{1}{2}\left[ \left( U_{1}\right) _{tt}+\left(
U_{1}\right) _{xx}\right] +i\delta (U_{2})_{t}+U_{2}+\left\vert
U_{1}\right\vert ^{2}U_{1} &=&i\Gamma U_{1},  \label{3} \\
i\left( U_{2}\right) _{z}+\frac{1}{2}\left[ \left( U_{2}\right) _{tt}+\left(
U_{2}\right) _{xx}\right] +i\delta (U_{1})_{t}+U_{1}+\left\vert
U_{2}\right\vert ^{2}U_{2} &=&-i\Gamma U_{2}.  \label{4}
\end{eqnarray}%
All the ingredients of the present system, including the separated gain and
loss \cite{PT-optics-exp}, can be realized experimentally in optics.
Nevertheless, the model based on Eqs. (\ref{3}) and (\ref{4}) turns out to
be irrelevant, because the linearized version of the equations for
excitations, in the usual form of $U_{1,2}\sim \exp \left( ikz-i\omega
t+iqx\right) $, \ gives rise to an unstable dispersion relation between the
propagation constant, $k$, real temporal frequency, $\omega $, and real
transverse wavenumber, $k$:
\begin{equation}
\left( k+\frac{1}{2}\omega ^{2}+\frac{1}{2}q^{2}\right) ^{2}=-\Gamma
^{2}+\left( 1+\delta \cdot \omega \right) ^{2}.  \label{k}
\end{equation}%
Indeed, it follows from Eq. (\ref{k}) that, in the presence of the gain and
loss, the zero solution is unstable, which is accounted for by an imaginary
part of $k$, against perturbations with $\left( 1+\delta \cdot \omega
\right) ^{2}<\Gamma ^{2}$. In other words, the $\mathcal{PT}$ symmetry is
always broken by the SOC terms in this system, while it gives rise to stable
2D solutions at $\Gamma =0$ \cite{opt-SOC-2D}.

A system which maintains the $\mathcal{PT}$ symmetry in the presence of the
SOC is based on the following coupled NLSEs:
\begin{eqnarray}
i\left( U_{1}\right) _{z}+\frac{1}{2}\left[ \left( U_{1}\right) _{tt}+\left(
U_{1}\right) _{xx}\right] -\delta \left( U_{2}\right) _{x}+U_{2}+\left\vert
U_{1}\right\vert ^{2}U_{1} &=&i\Gamma U_{1},  \label{5} \\
i\left( U_{2}\right) _{z}+\frac{1}{2}\left[ \left( U_{2}\right) _{tt}+\left(
U_{2}\right) _{xx}\right] +\delta \left( U_{1}\right) _{x}+U_{1}+\left\vert
U_{2}\right\vert ^{2}U_{2} &=&-i\Gamma U_{2}.  \label{6}
\end{eqnarray}%
This is a model of a planar coupler, in which the temporal dispersion of the
coupling coefficient is disregarded, while terms $\mp \delta \left(
U_{2,1}\right) _{x}$ account for ``skewness" of the coupling
in the transverse direction, assuming that the layer between the guiding
cores has an oblique structure. Roughly speaking, the latter means that
light tunnels between point $x$ in the first (second) core and point $%
x+\delta $ (respectively, $x-\delta $) in the mate core. The dispersion
relation for the linearized version of Eqs. (\ref{5}) and (\ref{6}) is%
\begin{equation}
\left( k+\frac{1}{2}\omega ^{2}+\frac{1}{2}q^{2}\right) ^{2}=1-\Gamma
^{2}+\left( \delta \cdot q\right) ^{2}.  \label{stable}
\end{equation}%
Unlike Eq. (\ref{k}), it demonstrates that the $\mathcal{PT}$ symmetry holds
at $\Gamma \leq 1$ (in fact, under the condition that the gain-loss
coefficient is smaller than the inter-core coupling constant, which is
scaled to be $1$).

The system may produce solitons in its \textit{bandgap}, i.e., at values of
the propagation constant, $k$, which cannot be obtained from Eq. (\ref%
{stable}). Simple analysis demonstrates that the dispersion relation (\ref%
{stable})\ gives rise to a \textit{semi-infinite bandgap}, in which solitons
are expected to exist:%
\begin{equation}
k>\left\{
\begin{array}{c}
\frac{1}{2}\left[ \delta ^{2}+\left( 1-\Gamma ^{2}\right) \delta ^{-2}\right]
,~\mathrm{at}~~\delta ^{2}>\sqrt{1-\Gamma ^{2}}, \\
\sqrt{1-\Gamma ^{2}},~\mathrm{at}~~\delta ^{2}<\sqrt{1-\Gamma ^{2}}.%
\end{array}%
\right.  \label{gap}
\end{equation}

\section{Solitons in the 1D system}

A family of stationary soliton solutions of the 1D version of Eqs. (\ref{5})
and (\ref{6}), in which the temporal dependence is dropped, are looked for
as
\begin{equation}
U_{1,2}(x,z)=\exp \left( ikz\right) u_{1,2}(x),  \label{Uu}
\end{equation}%
with complex functions $u_{1,2}(x)$ satisfying equations
\begin{eqnarray}
\frac{1}{2}u_{1}^{\prime \prime }-\delta u_{2}^{\prime }+u_{2}+\left\vert
u_{1}\right\vert ^{2}u_{1} &=&\left( k+i\Gamma \right) u_{1},  \label{1U} \\
\frac{1}{2}u_{2}^{\prime \prime }+\delta u_{1}^{\prime }+u_{1}+\left\vert
u_{2}\right\vert ^{2}u_{2} &=&\left( k-i\Gamma \right) u_{2},  \label{2U}
\end{eqnarray}%
with the prime standing for $d/dx$. The $\mathcal{PT}$ symmetry amounts to
the cross-symmetry constraint for the two components, which all the soliton
solutions obey:%
\begin{equation}
u_{1}(-x)=u_{2}^{\ast }(x)  \label{symm}
\end{equation}%
(the asterisk stands for the complex conjugation). The solutions are
characterized by their norm, i.e., the total power, in terms of the
underlying optics model:%
\begin{equation}
N=\int_{-\infty }^{+\infty }\left[ |u_{1}(x)|^{2}+|u_{2}(x)|^{2}\right] dx.
\label{norm}
\end{equation}

In the absence of the SOC terms ($\delta =0$), exact $\mathcal{PT}$%
-symmetric soliton solutions to Eqs. (\ref{1U}) and (\ref{2U}) are well
known \cite{Radik},
\begin{gather}
u_{2}(x)=e^{i\phi }\left( \sqrt{1-\Gamma ^{2}}+i\Gamma \right) u_{1}(x),~
\notag \\
u_{1}(x)=e^{i\phi }\sqrt{2\left( k-\sqrt{1-\Gamma ^{2}}\right) }\mathrm{sech}%
\left( \sqrt{2\left( k-\sqrt{1-\Gamma ^{2}}\right) }x\right) ,  \label{exact}
\end{gather}%
where $\phi $ is a constant phase. The exact stability condition for these
solutions is known too \cite{Radik}. Combined with the obvious existence
condition, $k>\sqrt{1-\Gamma ^{2}}$, it produces the following interval
filled by the stable $\mathcal{PT}$-symmetric solitons at $\delta =0$ (in
the absence of the SOC terms): $\sqrt{1-\Gamma ^{2}}<k<\left( 5/3\right)
\sqrt{1-\Gamma ^{2}}$. It is more relevant to write this in terms of the
norm of soliton (\ref{exact}), $N\left( \delta =0\right) =4\sqrt{2\left( k-%
\sqrt{1-\Gamma ^{2}}\right) }$:%
\begin{equation}
0<N<\left( 8/\sqrt{3}\right) \left( 1-\Gamma ^{2}\right) ^{1/4}\equiv N_{%
\mathrm{c}}\left( \delta =0\right) .  \label{interval}
\end{equation}

\begin{figure}[tbp]
\begin{center}
\includegraphics[height=4.cm]{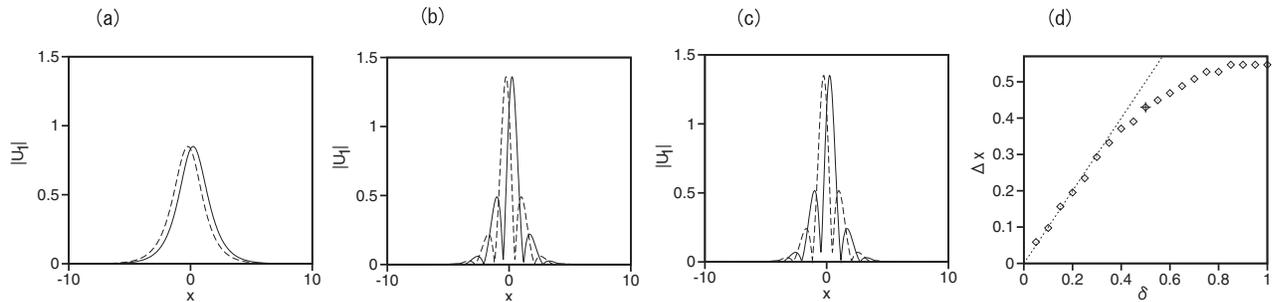}
\end{center}
\caption{Profiles of components $|u_{1}(x)|$ and $|u_{2}(x)|$ (solid and
dashed lines, respectively ) in the 1D solitons at the following values of
the SOC and gain-loss coefficients: (a) $\protect\delta =0.5$, $\Gamma =0.3$%
, (b) $\protect\delta =2$, $\Gamma =0.3$, and (c) $\protect\delta =2$, $%
\Gamma =0.8.$ All the solitons have the same total norm, $N=3$. (d) Distance
$\Delta x$ between peaks of the two components vs. $\protect\delta $ at $%
\Gamma =0.3$. }
\label{fig1}
\end{figure}

In the presence of the SOC ($\delta >0$), we have built stationary localized
solutions of the system by means of the well-known
imaginary-time-integration method \cite{Tosi}. Figures \ref{fig1}(a), (b),
and (c) show typical profiles of the stable solutions for $|u_{1}(x)|$ and $%
|u_{2}(x)|$, obtained at particular values of the SOC and $\mathcal{PT}$
parameters, $\delta $ and $\Gamma $, specified in the caption to Fig. \ref%
{fig1}. Selected for the comparison in this figure are the solitons which
all have equal norms, $N=3$.

A notable SOC effect is the splitting between peaks of the two components. A
simple perturbative analysis of Eqs. (\ref{1U}) and (\ref{2U}) demonstrates
that, for small $\delta $, the distance between the peaks is
\begin{equation}
\Delta x=\delta .  \label{Delta}
\end{equation}%
Figure \ref{fig1}(d) shows the numerically obtained distance $\Delta x$ as a
function of $\delta $ for $\Gamma =0.3$. Thus, relation (\ref{Delta}) is
virtually exact at $\delta \leq 0.4$.

It is worthy to note that the profiles of $\left\vert u_{1}(x)\right\vert $
and $\left\vert u_{2}(x)\right\vert $ are nearly identical for different
values of the gain-loss coefficient, $\Gamma =0.3$ and $0.8$, with the same $%
\delta =2$, in Figs. \ref{fig1}(b,c). This observation is in qualitative
agreement with the fact that, for given $N$, the profiles of exact solitons (%
\ref{exact}) do not depend on $\Gamma $ either.

An obvious effect of the increase of $\delta $ is a transition from the
smooth soliton profile to a multi-lobe structure, as seen from the
comparison of panels (a) and (b,c) in Fig. \ref{fig1}. This fact can be
explained by the inversion of the dispersion relation (\ref{stable}), to
express $q$ in terms of $k$, at $\omega =0$. The result of a simple algebra
is that, precisely at $k>\sqrt{1-\Gamma ^{2}}$, i.e., when the exact soliton
(\ref{exact}) exists for $\delta =0$, wavenumber $q$ becomes complex, which
implies the wavy profile of the soliton's tails, for
\begin{equation}
\delta ^{2}>k-\sqrt{k^{2}-\left( 1-\Gamma ^{2}\right) }.  \label{delta}
\end{equation}
The numerical results demonstrate that the transition occurs at $\delta
\approx 0.6$ for $\Gamma =0.3$, and at $\delta \approx 0.7$ for $\Gamma =0.8$%
, which is consistent with Eq. (\ref{delta}) in which numerically obtained
values of $k$ are substituted.

\begin{figure}[tbp]
\begin{center}
\includegraphics[height=4.cm]{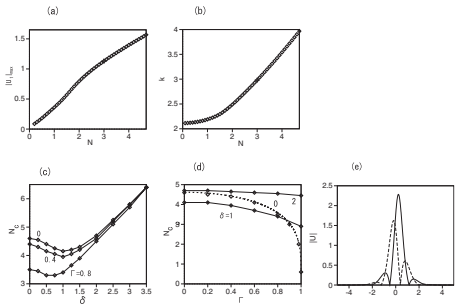}
\end{center}
\caption{(a) Maximum value of $|U_{1}|$ of stable 1D solitons, as a function
of $N$ for $\protect\delta =2$ and $\Gamma =0.3$. (b) The critical norm, $N_{%
\mathrm{c}}$ (above which stable 1D solitons do not exist), as a function of
the SOC strength, $\protect\delta $, for values of the gain-loss coefficient
$\Gamma =0$, $0.4$, and $0.8$. (c) The critical norm versus $\Gamma $, for $%
\protect\delta =0$, $1$, and $2$. (d) An example of a stable symmetric
solution found in the conservative model, with $\Gamma =0$, for $N=5$ and $%
\protect\delta =2$.}
\label{fig2}
\end{figure}

The stable soliton family is further characterized by Fig. \ref{fig2}(a),
which shows the largest value of the field $|u_{1}(x)|$ as a function of $N$%
, for $\delta =2$ and $\Gamma =3$. Naturally, $|u_{1}(x=0)|$ increases
monotonously with $N$, up to a critical value, $N_{\mathrm{c}}$, at which
the destabilization of the solitons happens via the breakup of the $\mathcal{%
PT}$ symmetry (the solitons exist at $N>N_{\mathrm{c}}$ as unstable
solutions, which cannot be found by means of the imaginary-integration
method). In the conservative system with $\Gamma =0$, stable asymmetric
solitons, with $u_{1}(x)\neq u_{2}(-x)$, may exist at $N>N_{\mathrm{c}}$,
but in the presence of $\Gamma >0$ this is impossible, as asymmetric
solitons would not maintain the balance between the gain and loss. Figure %
\ref{fig2}(d) shows an example of a stable asymmetric solution found at $%
\Gamma =0$ for $N=5$ and $\delta =2$

In fact, $N_{\mathrm{c}}$ is the most important characteristic of the
soliton families. It is shown, as a function of the SOC strength, $\delta $,
for fixed $\Gamma =0$, $0.4$, and $0.8$, in Fig. \ref{fig2}(b). In
particular, $N_{\mathrm{c}}$ is given by the exact expression (\ref{interval}%
) for $\delta =0.$ The initial decrease of $N_{\mathrm{c}}$, which is
observed in Fig. \ref{fig2}(b) at relatively small $\delta $, can be easily
understood: as shown by Eq. (\ref{Delta}), the two components get separated
by distance $\Delta x\approx \delta $, hence the effective linear coupling
between the components weakens with the increase of $\delta $, making a
smaller strength of the Kerr nonlinearity (measured by the norm) sufficient
to initiate the symmetry breaking between the two components. Obviously,
this effect should scale as $\delta ^{2}$, which is consistent with Fig. \ref%
{fig2}(b) at $\delta $ small enough. However, a nontrivial finding is that,
at $\delta $ exceeding values corresponding to the minimum of $N_{\mathrm{c}%
} $ in Fig. \ref{fig2}(b), $N_{\mathrm{c}}$ features \emph{rapid increase}
with $\delta $, which becomes asymptotically linear at large values of $%
\delta $. This property is explained in the next section which addresses the
limit form of system (\ref{5}), (\ref{6}) with the SOC terms dominating over
the paraxial diffraction.

Further, Fig. \ref{fig2}(c) shows $N_{\mathrm{c}}$ as a function of $\Gamma $
for fixed SOC strengths, $\delta =0$, $1$, and $2$, the dashed curve
representing the exact result given by Eq. (\ref{exact}) for $\delta =0$.
This figure shows too that the SOC terms tend to essentially expand the
stability area for the $\mathcal{PT}$-symmetric solitons, by increasing\ $N_{%
\mathrm{c}}$. However, even if $N_{\mathrm{c}}$ keeps a nonzero value up to $%
\Gamma =1$ at $\delta >0$, there is no stability region at $\Gamma >1$, as
the zero solution (the background of the solitons) is unstable in the latter
case, as follows from Eq. (\ref{stable}).

\section{The limit case of negligible intrinsic diffraction}

The results displayed above in Fig. \ref{fig2}(b) for large $\delta $
suggest to consider the limit case of the system in which the SOC terms
dominate over the paraxial diffraction. Dropping the second derivatives in
the 1D version of Eqs. (\ref{3}) and (\ref{4}), one thus arrives at the
simplified system,
\begin{eqnarray}
i\left( U_{1}\right) _{z}-\delta \left( U_{2}\right) _{x}+U_{2}+\left\vert
U_{1}\right\vert ^{2}U_{1} &=&i\Gamma U_{1},  \label{7} \\
i\left( U_{2}\right) _{z}+\delta \left( U_{1}\right) _{x}+U_{1}+\left\vert
U_{2}\right\vert ^{2}U_{2} &=&-i\Gamma U_{2}.  \label{8}
\end{eqnarray}%
whose dispersion relation is
\begin{equation}
k^{2}=1-\Gamma ^{2}+\left( \delta \cdot q\right) ^{2},  \label{beta2}
\end{equation}%
cf. Eq. (\ref{exact}). It gives rise to a finite bandgap, $k^{2}<$ $1-\Gamma
^{2}$ [unlike the semi-infinite bandgap given by Eq. (\ref{gap})], hence the
corresponding localized states may be considered as gap solitons. Actually,
the system of Eqs. (\ref{7}) and (\ref{8}) with $\Gamma =0$ is a new
conservative model generating gap solitons, therefore it makes sense to
consider its solutions, along with its $\mathcal{PT}$-symmetric version,
corresponding to $0<\Gamma <1$. An essential difference from the standard
gap solitons generated by the Bragg-grating model \cite{Bragg} is the
separation between peaks of the two components, and, on the other hand,
gap-soliton solutions are real in the present model.

Looking for stationary solutions to Eqs. (\ref{7}) and (\ref{8}) as per Eq. (%
\ref{Uu}), in the absence of the gain and loss, $\Gamma =0$, functions $%
u_{1}(x)$ and $u_{2}(x)$ satisfy a system of real equations%
\begin{eqnarray}
-ku_{1}-\delta u_{2}^{\prime }+u_{2}+u_{1}^{3} &=&0,  \label{U1} \\
-ku_{2}+\delta u_{1}^{\prime }+u_{1}+u_{2}^{3} &=&0,  \label{U2}
\end{eqnarray}%
whose solutions obey the cross-symmetry constraint, cf. Eq. (\ref{symm}): $%
u_{1}(-x)=u_{2}(x).$ In fact, $\delta $ may be easily absorbed into rescaled
coordinate $x$, therefore numerical results are presented below for $\delta
=1$. The total norm (\ref{norm}), defined according to the original
coordinate, scales as $\delta $, which explains the asymptotically linear $%
N_{\mathrm{c}}(\delta )$ dependence in Fig. \ref{fig2}(b).

It is possible to find exact soliton solutions to Eqs. (\ref{U1}) and (\ref%
{U2}), noting that the evolution of $u_{1}$ and $u_{2}$ along $x$ conserves
the corresponding formal Hamiltonian:%
\begin{equation}
h=\frac{k}{2}\left( u_{1}^{2}+u_{2}^{2}\right) -u_{1}u_{2}-\frac{1}{4}\left(
u_{1}^{2}+u_{2}^{2}\right) ^{2}+\frac{1}{2}u_{1}^{2}u_{2}^{2}.  \label{h}
\end{equation}%
Next, it is convenient to represent solutions in the ``polar" form,
\begin{equation}
u_{1}(x)=u(x)\cos \left( \theta (x)\right) ,~u_{2}(x)=u(x)\sin \left( \theta
(x)\right)  \label{UUU}
\end{equation}%
For soliton solutions which vanish at $x\rightarrow \pm \infty $, one should
set $h=0$, hence Eq. (\ref{h}) makes it possible to eliminate $u^{2}$ in
favor of $\theta $, after substituting expressions (\ref{UUU}):%
\begin{equation}
u^{2}=4\frac{k-\sin (2\theta )}{2-\sin ^{2}(2\theta )},  \label{U^2}
\end{equation}%
Finally, combining two equations (\ref{U1}) and (\ref{U2}) and Eq. (\ref{U^2}%
), it is easy to derive a single equation for $\theta (x)$:%
\begin{equation}
\delta \frac{d\theta }{dx}=k-\sin (2\theta ).  \label{theta}
\end{equation}

A solution\ to Eq. (\ref{theta}), which generates a soliton after the
substitution in Eq. (\ref{U^2}), exists for $k^{2}<1$:
\begin{equation}
\theta =-\frac{\pi }{4}+\arctan \left[ \frac{1+k}{\sqrt{1-k^{2}}}\tanh
\left( \frac{\sqrt{1-k^{2}}}{\delta }x\right) \right] .  \label{<0}
\end{equation}%
The total norm of the solitons calculated as per Eqs. (\ref{norm}) and (\ref%
{UUU})-(\ref{<0}), can be written as%
\begin{equation}
N(k)=2\sqrt{2}\delta \left\{
\begin{array}{c}
-\arctan \left( \sqrt{2\left( 1-k^{2}\right) }/k\right) ,~-1<k<0, \\
\left[ \pi -\arctan \left( \sqrt{2\left( 1-k^{2}\right) }/k\right) \right]
,~0<k<1.%
\end{array}%
\right.  \label{N}
\end{equation}%
Thus, with the increase of $k$ from $-1$ to $+1$, $N(k)$ monotonously grows
from $N(k=-1)=0$ to $N(k=+1)=2\sqrt{2}\pi \delta $. The analytical solution
makes it possible to find the distance between peaks of the two components,
cf. Eq. (\ref{Delta}). In the general case, the analytical expression for $%
\Delta x$ is cumbersome. It takes a relatively simple form for $k=0$, which
corresponds to a stable gap soliton (see below):%
\begin{equation}
\Delta x(k=0)=2\delta \mathrm{Artanh}\left( \frac{1-3^{1/4}\left( \sqrt{3}%
-1\right) }{2-\sqrt{3}}\right) \approx 0.275~\delta .
\end{equation}

\begin{figure}[tbp]
\begin{center}
\includegraphics[height=4.cm]{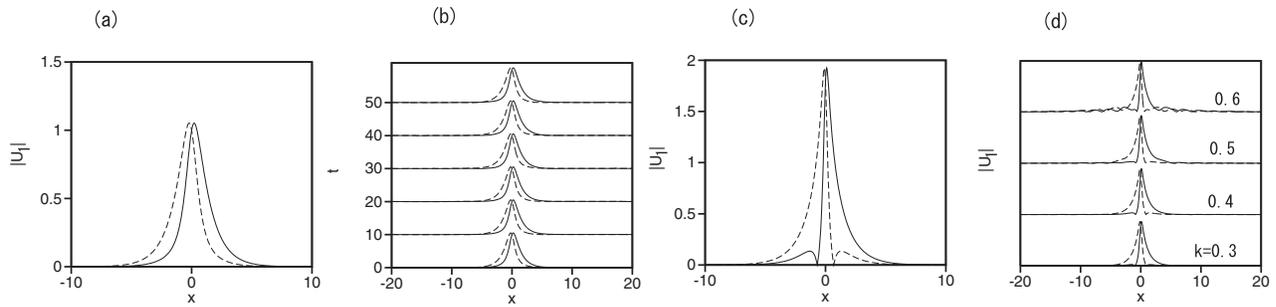}
\end{center}
\caption{(a) The exact solution for $|u_{1}(x)|$ and $|u_{2}(x)|$ (solid and
dashed lines), given by Eqs. (\protect\ref{UUU})-(\protect\ref{N}) for $%
k=-0.5$ and $\protect\delta =1$. (b) The perturbed evolution of component $%
\left\vert U_{1}(x,t)\right\vert $ of the same gap soliton, corroborating
its stability. (c) The exact solution for $k=0.6$ and $\protect\delta =1$.
(d) Snapshots produced by the simulations of the perturbed evolution of the
gap solitons with $k=0.3,0.4,0.5$, and $0.6$, at $z=800$. The results reveal
a boundary between stable and unstable solitons at $k\approx 0.5$.}
\label{fig3}
\end{figure}

Figure \ref{fig3}(a) displays an example of the exact solution for $k=-0.5$
and $\delta =1$, whose total norm is $N=3.35$, in agreement with Eq. (\ref{N}%
). Further, Fig. \ref{fig3}(b) shows a result of the test of the stability
of this gap soliton, produced by direct simulations of Eqs.~(\ref{7}) and (%
\ref{8}) with $\Gamma =0$ and small random perturbations added to the
initial conditions. Figure \ref{fig3}(b) clearly shows that the gap soliton
is stable.

Figure \ref{fig3}(c) displays another example of the exact soliton,for $%
k=0.6 $ and $\delta =1$, with the total norm $N=5.82$, which also agrees
with Eq. (\ref{N}). In this case, the simulations, the result of which is
shown in the top plot of Fig. \ref{fig3}(d) at $z=800$, demonstrate weak
instability of the soliton, which leads to generation of an undulating tail.
The simulations for other values of $k$, that are displayed too in Fig. \ref%
{fig3}(d), suggest that the intrinsic boundary between stable and unstable
gap solitons is located at $k\approx 0.5$, cf. a qualitatively similar
results for the gap solitons in the Bragg-grating model \cite%
{Bragg-stability}.

\begin{figure}[tbp]
\begin{center}
\includegraphics[height=4.cm]{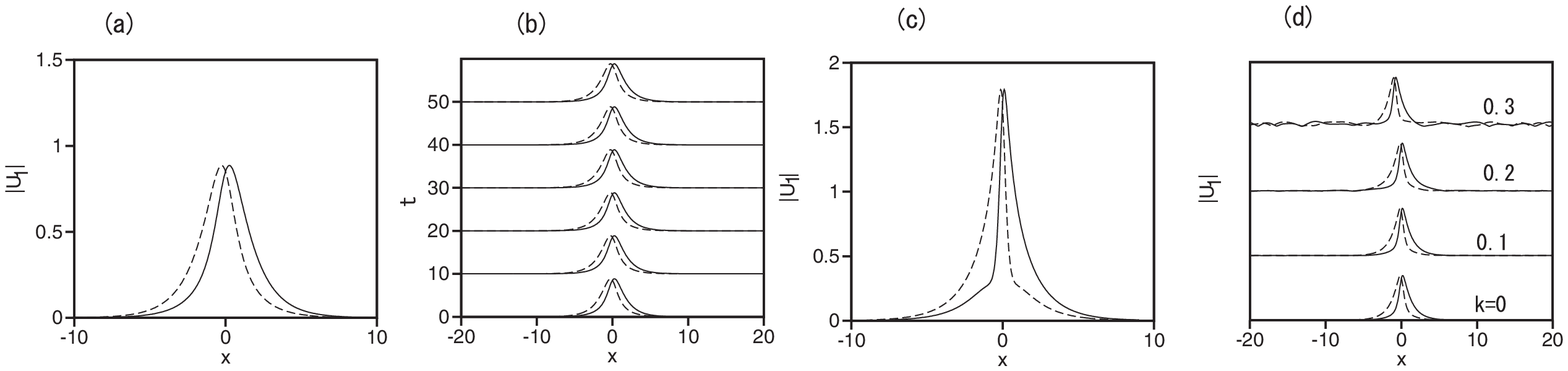}
\end{center}
\caption{(a) Numerically obtained stationary solutions $|u_{1}(x)|$ and $%
|u_{2}(x)|$ (solid and dashed lines) of Eqs. (\protect\ref{7} and (\protect
\ref{8}) with $\Gamma =0.5$ and $\protect\delta =1$ (the $\mathcal{PT}$%
-symmetric version of the system in which the intrinsic diffraction is
omitted) at $k=-0.5$. (b) The perturbed evolution of component $\left\vert
U_{1}\left( x,t\right) \right\vert $ of the same soliton.(c) The numerically
obtained stationary soliton with $k=0.6$. (d) Results of the perturbed
evolution of the solitons with $k=0$, $0.1$, $0.2$, and $0$ (all for $\Gamma
=0.5$) at $z=900$, which reveal the location of the stability boundary at $%
k\approx 0.2$.}
\label{fig4}
\end{figure}

The $\mathcal{PT}$-symmetric version of Eqs. (\ref{7}) and (\ref{8}) with $%
\Gamma >0$ was solved numerically, which also produced solitons. Two
examples, for $\delta =1$ and $\Gamma =0.5$, but different propagation
constants, $k=-0.5$ and $k=0.6$, with the norms, respectively, $N=3.20$ and $%
6.31$, are displayed in Figs. \ref{fig4}(a) and (c). Simulations of the
perturbed evolution of the former soliton demonstrate its stability in Fig. %
\ref{fig4}(b). On the other hand, a set of results of the simulations of the
solitons with several values of $k$, presented in Fig. \ref{fig4}(d) at $%
z=900$, make it evident that the latter soliton is unstable, the stability
boundary being located at $k\approx 0.2$.

\section{Two-dimensional solitons}

\begin{figure}[tbp]
\begin{center}
\includegraphics[height=4.cm]{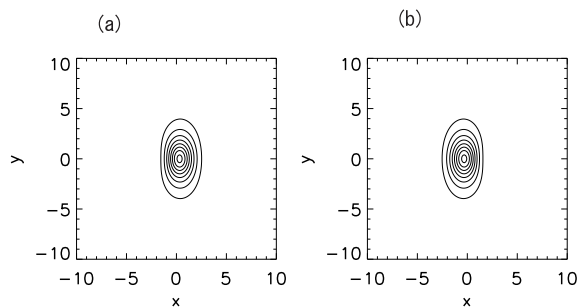}
\end{center}
\caption{Contour plots of fields of fields $|u_{1}\left( x,t\right) |$ (a)
and $|u_{2}(x,t)|$ (b) in a numerically found 2D soliton, at $\protect\delta %
=1,\Gamma =0.2$, with norm $N=7.5$.}
\label{fig5}
\end{figure}

\begin{figure}[tbp]
\begin{center}
\includegraphics[height=6.cm]{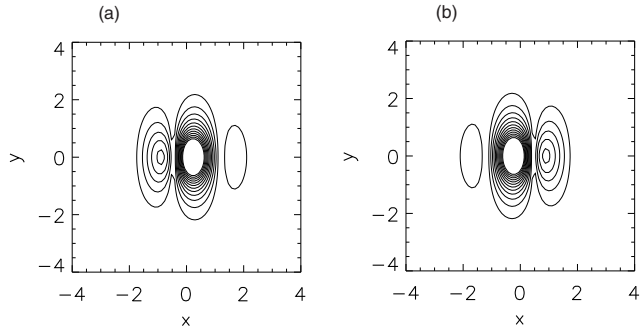}
\end{center}
\caption{The same as in Fig. \protect\ref{fig5}, but for $\protect\delta %
=2,\Gamma =0.2$, and $N=7.1$.}
\label{fig6}
\end{figure}

Numerical solution of the full 2D system of Eqs. (\ref{3}) and (\ref{4})
produces stable solitons too, which are characterized by the respective
norm,
\begin{equation}
N=\int \int dxdt\left[ \left\vert U_{1}(x,t)\right\vert ^{2}+\left\vert
U_{2}\left( x,t\right) \right\vert ^{2}\right] .  \label{N2D}
\end{equation}%
Examples of the 2D solitons are displayed in Figs. \ref{fig5} and \ref{fig6}%
, which, in particular, demonstrate the mirror symmetry of $\left\vert
U_{1}\left( x,t\right) \right\vert $ and $\left\vert U_{2}\left( x,t\right)
\right\vert $. Note that, similar to what was observed above for 1D
solitons, the increase of the SOC strength, $\delta $, generates a complex
multi-lobe shape of the 2D solitons.

\begin{figure}[tbp]
\begin{center}
\includegraphics[height=4.cm]{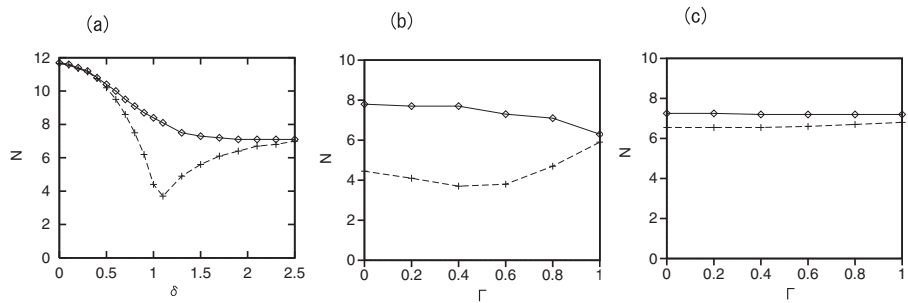}
\end{center}
\caption{(a) The upper and lower critical values of norm (\protect\ref{N2D})
for the 2D solitons, as functions of the SOC strength $\protect\delta $, for
a fixed value of the gain-loss coefficient, $\Gamma =0.1$. (b) and (c) The
critical values as functions of $\Gamma $ for $\protect\delta =1$ and $%
\protect\delta =2$, respectively. Stable 2D solitons exist between the
continous and dashed critical lines. }
\label{fig7}
\end{figure}

As shown in Fig. \ref{fig7}, two different critical values of the norm can
be identified for the 2D solitons. The upper one, $N_{\mathrm{c}}^{\mathrm{%
(upp)}}$, is the value at which, similar to what is found above for the 1D
system, the $\mathcal{PT}$-symmetric solitons are destabilized by the
spontaneous symmetry breaking. Specific to the 2D setting is a lower limit, $%
N_{\mathrm{c}}^{\mathrm{(low)}}$, below which the solitons cannot self-trap.
At $N<$ $N_{\mathrm{c}}^{\mathrm{(low)}}$ simulations demonstrate spreading
of the wave fields.

The existence of stable 2D solitons in the free space, under the action of
the cubic-only attractive nonlinearity, is a nontrivial finding, as it was
commonly believed that all solitons in such setting are destabilized by the
(critical) collapse \cite{collapse,Bilbao}. Recently \cite{we}, it was reported
that the linear SOC terms may stabilize 2D solitons, as these terms break
the specific scaling invariance of the 2D\ NLSEs, which accounts for the
critical collapse in this case. The stability regions of the $\mathcal{PT}$%
-symmetric 2D solitons, demonstrated in Fig. \ref{fig7}, are a still
stronger result, as the solitons are able to stay stable despite the
simultaneous presence of two major destabilization factors: the possibility
of the collapse, and the trend to spontaneous breakup of the balance between
the gain and loss in the two cores of the coupler. It is pertinent to
mention that other types of the SOC terms, which are relevant to
two-component BEC, do not give rise to $N_{\mathrm{c}}^{\mathrm{(low)}}$,
the respective stability region being $0<N<N_{\mathrm{c}}^{\mathrm{(upp)}}$
\cite{we}. However, the $\mathcal{PT}$-symmetry cannot be introduced in a
physically relevant form in that setting.

Naturally, Fig. \ref{fig7}(a) shows that the stability region found in the
present system, $N_{\mathrm{c}}^{\mathrm{(low)}}<N<N_{\mathrm{c}}^{\mathrm{%
(upp)}}$, shrinks to nothing in the limit of $\delta \rightarrow 0$, when
the 2D modes degenerate into the commonly known unstable \textit{Townes
solitons} \cite{collapse}. On the other hand, a nontrivial finding is that
the stability region attains its largest size at a finite value of the SOC
strength, $\delta \sim 1$, as seen in Fig. \ref{fig7}. Similar to the 1D
system [cf. Fig. \ref{fig2}(c)], the stability region keeps a (small) finite
width at $\Gamma =1$, completely disappearing at $\Gamma >1$.

\section{Conclusion}

The objective of this work is to introduce 1D and 2D optical systems which
makes it possible to emulate the $\mathcal{PT}$ symmetry in combination with
effects of the SOC\ (spin-orbit coupling), thus creating a novel physical
setting. The systems are based on dual-core planar optical couplers with the
Kerr nonlinearity in their cores. The $\mathcal{PT}$ symmetry is represented
by equal amounts of the linear gain and loss in the two cores, while the SOC
is induced by the skew form of the coupling between the cores. Families of
stable 1D and 2D $\mathcal{PT}$-symmetric solitons have been identified by
means of numerical and analytical methods. The size of the stability region
of the 1D solitons nonmonotonously depends of the SOC strength, $\delta $,
originally shrinking and then rapidly expanding with the increase of $\delta
$. In the limit of the SOC terms dominating over the paraxial diffraction,
the 1D system produces a new model for gap solitons, for which exact
analytical solutions have been found. 2D solitons may be stable against the
combined action of two major destabilization factors, \textit{viz}., the
critical collapse driven by the Kerr self-focusing, and the trend to
spontaneous breakup of the balance between the gain and loss in the coupled
cores.

The present analysis may be extended in other directions. In particular, it
is interesting to consider the mobility of 1D and 2D solitons and collisions
between them, which is a nontrivial problem in the presence of the SOC \cite%
{we}. Dynamical regimes, such as Josephson-like oscillations of localized
modes between the coupled cores, may be interesting too.

\end{document}